\documentclass[acmtog]{acmart}
\copyrightyear{2026}
\acmYear{2026}
\setcopyright{cc}
\setcctype{by}
\acmConference[SA Conference Papers '26]{SIGGRAPH Asia 2026 Conference Papers}{December 01--04, 2026}{Kuala Lumpur, Malaysia}
\acmBooktitle{SIGGRAPH Asia 2026 Conference Papers (SA Conference Papers '26), December 01--04, 2026, Kuala Lumpur, Malaysia}
\acmDOI{10.1145/3829340.3842202}
\acmISBN{979-8-4007-2842-6/2026/12}

\usepackage{hyperref}
\usepackage{xr}
\graphicspath{{./images/}}

\usepackage{colortbl}    % \cellcolor{}
\usepackage{xcolor}      % named colors

\definecolor{rankone}{HTML}{F78A8A}     % strongest pink/red
\definecolor{ranktwo}{HTML}{F8B68A}     % orange
\definecolor{rankthree}{HTML}{FBE38B}   % yellow

\usepackage{soul}

\usepackage[linesnumbered]{algorithm2e}

\newif\ifsubmit
\submittrue
\ifsubmit
    
    \newcommand{\TODO}[1]{}

    \newcommand{\rev}[1]{#1}
    
    \newcommand{\revbegin}{}
    \newcommand{\revend}{}

    \newcommand{\deleted}[1]{}

    \newcommand{\sayit}[3]{}

    \newcommand{\yotam}[2][1=]{}
    \newcommand{\ted}[2][1=]{}
    \newcommand{\jason}[2][1=]{}
    \newcommand{\jianchao}[2][1=]{}
    \newcommand{\jose}[2][1=]{}

    \newcommand{\yot}[1]{}
\else
    \definecolor{todo}{rgb}{0.99,0.59,0.55}
    \newcommand{\TODO}[1]{\textcolor{red}{\textbf{*** #1 ***}}}
    
    \newcommand{\rev}[1]{\textcolor{blue}{#1}}
    
    \newcommand{\revbegin}[1]{\color{blue}} % \revbegin
    \newcommand{\revend}[1]{\color{black}} % \revend

    \newcommand{\deleted}[1]{\st{#1}}

    \newcommand{\sayit}[3]{{\small\protect\colorlet{col}{#2}\color{col}\colorbox{col!15}{\textsc{#1:}} #3}}
    
    \definecolor{cyan3}{HTML}{C99E10} %
    \usepackage[colorinlistoftodos,prependcaption,textsize=tiny,color=todo,tickmarkheight=0.1cm]{todonotes}
    \setuptodonotes{noinline}
    \definecolor{lime}{rgb}{0.53,0.78,0.27}
    \newcommandx{\yotam}[2][1=]{\todo[linecolor=lime,backgroundcolor=lime!25,bordercolor=lime,#1]{\colorbox{lime!50}{\textsc{Yotam:}} #2}}
    \newcommandx{\ted}[2][1=]{\todo[linecolor=teal,backgroundcolor=teal!25,bordercolor=teal,#1]{\colorbox{teal!50}{\textsc{Ted:}} #2}}
    \newcommandx{\jason}[2][1=]{\todo[linecolor=orange,backgroundcolor=orange!25,bordercolor=orange,#1]{\colorbox{orange!50}{\textsc{Jason:}} #2}}
    \newcommandx{\jianchao}[2][1=]{\todo[linecolor=blue,backgroundcolor=blue!25,bordercolor=blue,#1]{\colorbox{blue!50}{\textsc{Jianchao:}} #2}}
    \newcommandx{\jose}[2][1=]{\todo[linecolor=magenta,backgroundcolor=magenta!25,bordercolor=magenta,#1]{\colorbox{magenta!50}{\textsc{Jose:}} #2}}

    \newcommand{\yot}[1]{\sayit{Yotam}{lime}{#1}}
    
\fi

\usepackage{xspace}

\definecolor{quote-gray}{gray}{0.3}

\begin{document}

\newcommand{\system}{\emph{ColorGradedGaussians}\xspace}
\title{Reparametrizing 3D Gaussian Splatting for Real-Time Palette-based Color and Luminance Editing}
% Palette-Based Color, Curve, and Pixelwise Color Grading for 3D Gaussian Splatting with View-space Sparse Decomposition

% dummy placeholder for anonymous and page length

\author{Cheng-Kang Ted Chao}
\email{cchao8@gmu.edu}
\affiliation{%
  \institution{George Mason University}
  \country{USA}
}

\author{Yotam Gingold}
\email{ygingold@gmu.edu}
\affiliation{%
  \institution{George Mason University}
  \country{USA}
}

\renewcommand{\shortauthors}{Cheng-Kang Ted Chao and Yotam Gingold}

\begin{teaserfigure}%
\centering
\includegraphics[width=\textwidth]{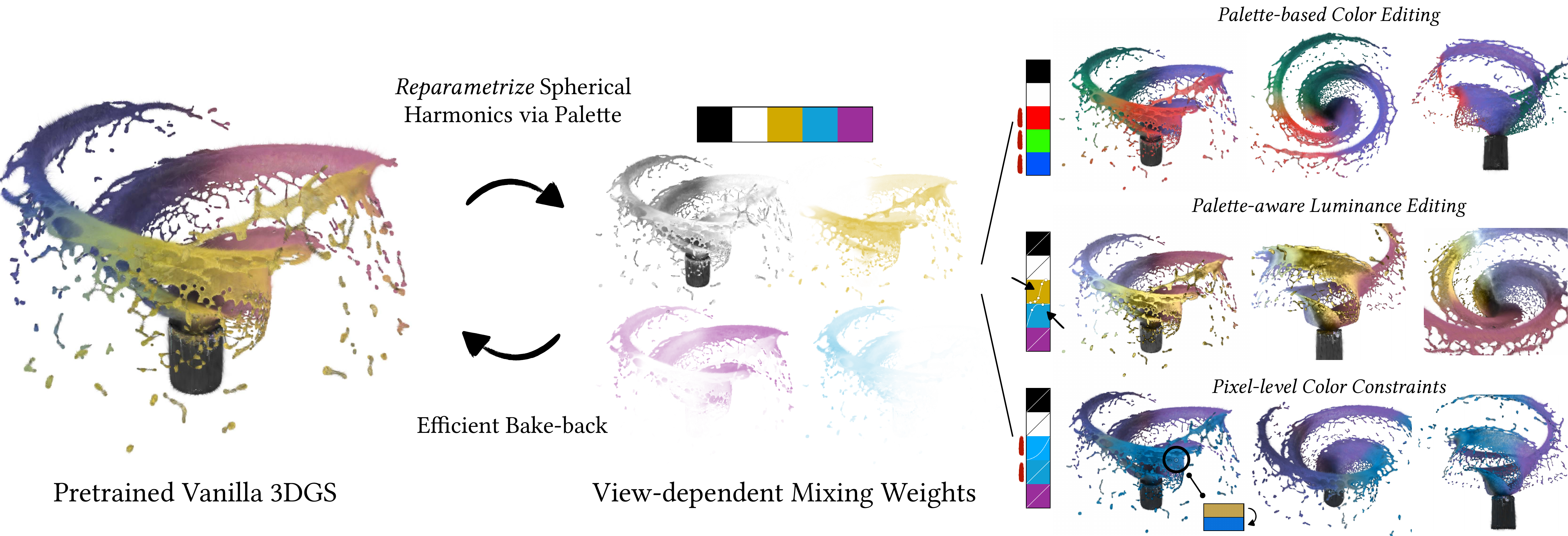}
\caption{
    Given a pretrained vanilla 3DGS \emph{(left)}, our method reparametrizes its per-Gaussian color spherical harmonics (SH) into view-dependent palette-weight SH \emph{(middle; white component omitted since the scene's background is white)}, yielding sparse, high-quality weights for editing. The user edits the scene through three real-time interactions \emph{(right)}: palette manipulation, per-palette tone curves, and pixel-level color constraints (the click marker indicates a constraint from dark yellow to blue), all propagating consistently across novel views. A constraint-driven solver jointly satisfies the three modalities in tens of milliseconds, and the edited representation bakes back into a vanilla 3DGS for standard viewers: palette edits at no additional cost, tone-curve edits in a few seconds. \emph{Scene:} \href{https://superspl.at/scene/56a89ca0}{``SPLAT PAINT -- EXPERIMENT''} by \href{https://superspl.at/user?id=sa3d}{St\'ephane Agullo}.
}
\label{fig:teaser}
\Description{}
\end{teaserfigure}

\begin{abstract}
Professional color editing requires precise control over both color (hue and saturation) and lightness, ideally through separate, independent controls. We present a real-time interactive color editing framework for 3D Gaussian Splatting that supports palette-based recoloring, per-palette tone curves for color-aware luminance adjustment, and pixel-level color constraints. Rather than training a new representation from scratch, we reparameterize the spherical harmonics of a pretrained vanilla 3DGS to encode view-dependent palette weights. We simultaneously solve for weights and palette colors via a loss based on image-space sparsity. Luminance editing is realized as a per-pixel weight shift along the achromatic axis, which we show is equivalent to a per-pixel palette-aware luminance edit. This view-space formulation addresses a core limitation of prior primitive-space methods, where alpha-blending breaks per-Gaussian sparsity and causes edits to bleed into unintended regions. Our edits run in tens of milliseconds via an iteratively reweighted least squares and damped block-coordinate descent that couples tone curves and palette shifts under view-space sparsity. Our representation can be efficiently baked back into a vanilla 3DGS, preserving compatibility with standard viewers. We demonstrate sparser, more localized edits than prior palette-based 3DGS methods, while enabling independent luminance control per palette color and view-consistent pixel-level constraints, capabilities previously unavailable for 3DGS.
\end{abstract}




\maketitle

\begin{figure*}[!t]
\includegraphics[width=\linewidth,scale=1]{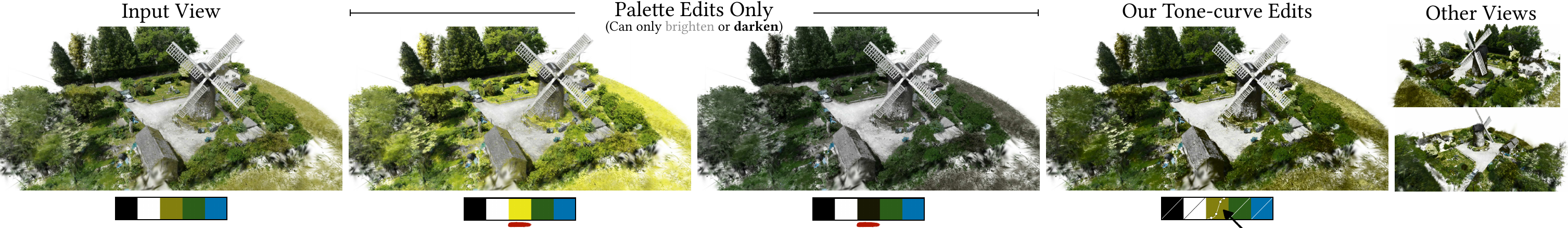}
\caption{
\textbf{Palette-based tone-curve edits in RGB-space.} Given the input view \emph{(left)}, shifting the green palette color along its luminance axis uniformly brightens \emph{(second)} or darkens \emph{(third)} every pixel that uses it. Our per-palette tone curve \emph{(fourth)} reshapes the luminance mapping for green independently, adding contrast across the greens (trees and grass) while leaving other colors untouched, and propagates consistently across novel views \emph{(right)}. We formulate tone curves in RGB-space rather than Lab-space~\cite{chao2023colorfulcurves} to keep palette bake-back closed-form. \emph{Scene:} \href{https://superspl.at/scene/401cdb2a}{"Berkswell Windmill"} by \href{https://superspl.at/user?id=ijenko}{Ian Jenkins}.
}
\label{fig:rgb-tone-curves}
\end{figure*}

\section{Introduction}
Color grading for 3D scenes is central to modern content creation, from visual effects and film production~\cite{debevec2008rendering} to virtual production and augmented reality~\cite{jeffrey2020ves}. 
Beyond global adjustments, artists need selective control over specific color ranges, with edits that remain photorealistic and geometrically consistent across viewpoints. 
In 2D, palette-based editing~\cite{chang2015palette, Tan:RGB2016, tan2018efficient} offers an intuitive workflow by expressing each pixel as a mixture of a small set of palette colors and propagating edits by keeping the mixing weights fixed. 
This paradigm has been improved through better palette extraction and sparser, more localized decompositions~\cite{aksoy2017unmixing, wang2019improved, chao2023locopalettes, zhang2025semantic}. Professional workflows benefit from independent lightness control per color range, e.g., via palette-aware tone curves~\cite{chao2023colorfulcurves}. 
Extending these capabilities to 3D scenes requires not only multi-view consistency, but also palette decompositions that remain sparse and editable in the rendered images users observe and edit.

Recent reconstruction methods such as NeRF~\cite{mildenhall2021nerf} and 3D Gaussian Splatting (3DGS)~\cite{kerbl20233d} enable photorealistic novel-view synthesis, and several methods adapt palette-based editing to NeRF~\cite{kuang2023palettenerf, gong2023recolornerf, wu2022palettenerf} and 3DGS~\cite{PaletteGaussian, RecolorGaussian} by decomposing each primitive's diffuse color into weighted combinations of palette colors.
However, primitive-level decomposition suffers from a fundamental mismatch: alpha-blending many Gaussians at a single pixel breaks per-primitive sparsity, producing dense pixel-level decompositions that cause palette edits to bleed into unintended regions \rev{(Fig.~\ref{fig:sparsity-compare})}.
Palette-based NeRF methods face similar issues and additionally require expensive preprocessing for weight guidance.
Prior palette-based 3DGS methods also lack color-aware tone curves, and their pixel-level color constraints require joint re-optimization of palette and view-dependent components, preventing real-time interaction (see Fig.~6 in supplemental)

We present a real-time color grading framework for 3DGS that resolves these limitations by finetuning a pretrained vanilla 3DGS into a \emph{palette subspace}, where each Gaussian's spherical harmonics (SH) encode view-dependent palette weights instead of RGB colors. Palette decomposition then occurs at the pixel-level after splatting, in the same view-space where edits are perceived and specified. To absorb the storage overhead, we factorize the higher-order weight SH into rank-1 form with little quality loss (Table~\ref{tab:nvs_comparison}). This view-space formulation supports three complementary editing controls (Fig.~\ref{fig:teaser},~\ref{fig:rgb-tone-curves}, and~\ref{fig:gallery1}): direct palette manipulation, per-palette tone curves for color-aware luminance adjustment, and pixel-level color constraints from any rendered view. To ensure perceptually similar colors receive similar decompositions, we introduce a color consistency loss based on inverse barycentric coordinates with respect to the current palette. We adapt the 2D constraint-driven solver of \citet{chao2023colorfulcurves} via iteratively reweighted least squares (IRLS) and damped block-coordinate descent (dBCD) that couples tone curves and palette shifts, running in tens of milliseconds per edit. Edits bake back into a vanilla 3DGS at essentially no cost for palette edits, or with a brief finetuning step (a few seconds) for luminance edits, preserving compatibility with standard viewers. We evaluate on multiple scenes and show that our framework provides editing capabilities unavailable in prior palette-based 3DGS methods while also improving edit quality and interactivity. \rev{Code for this work can be found at \href{https://github.com/tedchao/ReparamGS-Palette}{https://github.com/tedchao/ReparamGS-Palette}.}

\section{Related Work}
\paragraph{Appearance editing in neural scene representations.}
Neural scene representations such as NeRF~\cite{mildenhall2021nerf} and 3DGS \cite{kerbl20233d} have enabled a wide range of appearance editing techniques by explicitly modeling view-dependent radiance. Text-driven editing~\cite{haque2023instruct, zhuang2023dreameditor, chen2024gaussianeditor, kamata2023instruct, bao2023sine} and neural style transfer~\cite{zhang2022arf, gu2021stylenerf, liu2024stylegaussian, huang2022stylizednerf, fan2022unified, zhang2025stylizedgs} support semantic or artistic modifications but lack direct color control and often struggle with multi-view consistency, making them unsuitable for professional color grading. Relighting and tone mapping methods~\cite{gao2024relightable, cui2025luminance, liu2025gausshdr} target photometric consistency through physically based rendering or luminance-domain processing, but offer no palette-based abstractions or selective color control. Segmentation-driven editing methods~\cite{cen2023saga, chenkarf, tsang2026artisangs, rutayisire2025recogs, mazzucchelli2026virgi} support object- or region-level recoloring, and intrinsic decomposition methods~\cite{10.1145/3804495} enable material-level edits via per-pixel reflectance-shading separation.
% \yotam{Intrinsic decomposition allow for separate control over scene, lighting, and per pixel color, but not color coherent editing (or per color tone curves?).}
These approaches decompose scenes by spatial regions or material components rather than by color, targeting a different class of edits: they do not provide per-color tone-curve control, color-coherent propagation across spatially disconnected regions, or interactive constraint-driven optimization. They are orthogonal to our palette-based formulation.

\begin{figure*}[t]
\includegraphics[width=\linewidth]{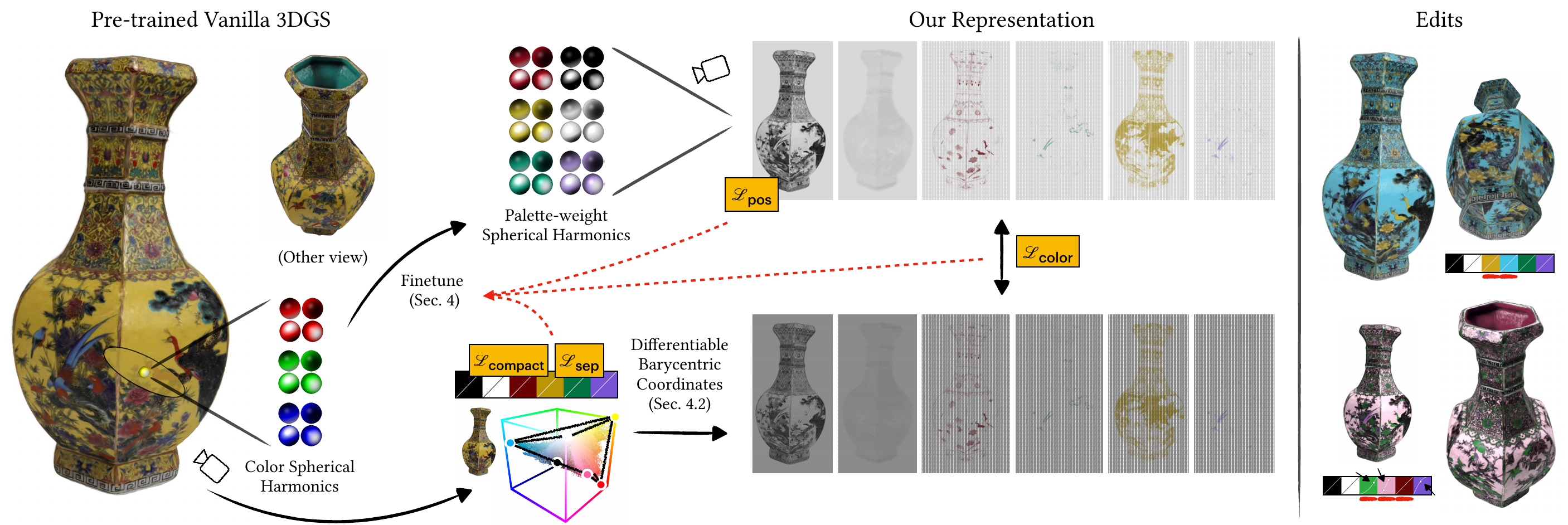}
\caption{
    \rev{
    \textbf{Method overview.} Starting from a vanilla 3DGS (\emph{left}), we reparametrize the scene into a palette subspace. We freeze all geometric parameters (position, covariance, opacity) and finetune each splat's color spherical harmonics (SH) into palette-weight SH, jointly optimized with the palette (Sec.~\ref{sec:optimization}). Our reparametrization splats palette weights rather than colors, shown in the middle as per-palette weight layers from a given view. We supervise these view-space splatted weights against the barycentric decomposition of the frozen splatted color over the current palette (Sec.~\ref{sec:barycentric_supervision}), which makes the weights sparse and assigns similar colors consistent weights. Red dotted lines denote the gradients of our losses ($\mathcal{L}_\text{color}$, $\mathcal{L}_\text{pos}$, $\mathcal{L}_\text{compact}$, $\mathcal{L}_\text{sep}$) backpropagated for finetuning. The resulting representation supports palette-based recoloring and luminance edits consistent across views (\emph{right}).
    \emph{Scene:} \href{https://superspl.at/scene/74c6d395}{``Container''} by \href{https://superspl.at/user/yyw}{Wu (Creality PIKA)}.
    }
}
\label{fig:pipeline}
\end{figure*}

Closer to our work, \citet{tojo2022recolorable} extends posterization-style editing~\cite{chao2021posterchild} to NeRFs for synthetic scenes, while PaletteNeRF~\cite{kuang2023palettenerf} and RecolorNeRF~\cite{gong2023recolornerf} support real-world scenes via palette decomposition of volumetric points, at the cost of expensive preprocessing and slow rendering.
Recent palette-based 3DGS methods~\cite{PaletteGaussian, RecolorGaussian} achieve real-time performance by decomposing each Gaussian's color into mixtures of global palette colors, but their primitive-level decomposition does not survive alpha-blending: sparsity at the primitive level does not translate into sparsity in the rendered image, manifesting as color bleeding (\rev{Fig.~5 in supplemental}).
We address this limitation by reparametrizing a pretrained 3DGS into a palette subspace where decomposition occurs at the pixel level after splatting, while remaining compatible with standard 3DGS viewers via bake-back.

\paragraph{Palette-based image recoloring.}
Palette-based image recoloring enables intuitive color control by representing each pixel as a mixture of a small set of representative palette colors. \citet{chang2015palette} introduced this paradigm via color clustering and radial-basis-function-weighted color deformation. Subsequent work formulated palette extraction and decomposition geometrically in color space: \citet{Tan:RGB2016} proposed convex-hull-based decomposition into translucent color layers, and \citet{tan2018efficient} extended this to spatially coherent additive mixing weights in RGBXY-space, enabling recoloring and color harmonization~\cite{tan2025palette}. Other approaches improve palette quality via optimization~\cite{wang2019improved} or coarse-to-fine convex hull construction~\cite{sun2023building}. Continuous palette formulations~\cite{shugrina2020nonlinear} provide an alternative to discrete palettes for direct manipulation of color distributions.
Independent of palette extraction, \citet{chao2023colorfulcurves} introduced palette-aware tone curves to support independent lightness and contrast control per palette color and image-space color constraints.
Extensions to video recoloring compute temporally coherent RGBT palettes~\cite{Du:2021:VRS}, with later work accelerating palette evolution via Bézier interpolation between frames~\cite{du2025fast}.
A fundamental limitation of palette-based recoloring is its difficulty in selectively editing objects that share similar colors, addressed by incorporating segmentation~\cite{chao2023locopalettes} or by extracting semantic palettes in high-dimensional feature spaces~\cite{du2024palette, zhang2025semantic}. Unmixing approaches~\cite{aksoy2016interactive, aksoy2017unmixing, koyama2018decomposing} instead formulate recoloring as energy minimization to produce sparse layers of nearly homogeneous colors. All these methods operate in image space, where palette weights are defined, regularized, and edited at the pixel-level; this assumption breaks down for 3D scenes (Fig.~4 in supplemental), motivating our view-space formulation.

\section{Background and Motivation}
\label{sec:motivation}
Recent work~\cite{PaletteGaussian, RecolorGaussian} extends 3DGS~\cite{kerbl20233d} to palette-based color editing by decomposing each Gaussian's diffuse color as a linear combination of a small set of palette colors. Let $K$ denote the number of palette colors, $\mathbf{P} \in \mathbb{R}^{K \times 3}$ a trainable global palette shared across the scene, and $\mathbf{w}_i \in \mathbb{R}^{K}$ a per-Gaussian weight vector with $w_{i,k} \in [0,1]$ and $\sum_{k} w_{i,k} = 1$. Gaussian $i$'s diffuse component is then $\mathbf{c}_i^d = \mathbf{w}_i^{\top} \mathbf{P}$, and view-dependent effects are captured by an additive specular component $\mathbf{c}_i^s(\theta, \phi)$ at viewpoint $(\theta, \phi)$, modeled via low-order spherical harmonics~\cite{PaletteGaussian} or neural offsets~\cite{RecolorGaussian}, yielding the full per-Gaussian color $\mathbf{c}_i(\theta, \phi) = \mathbf{c}_i^d + \mathbf{c}_i^s(\theta, \phi)$. Weights $\mathbf{w}_i$ are stored per-Gaussian or, for compression, decoded from each Gaussian's position through a small Hash-MLP~\cite{muller2022instant}, and jointly optimized with $\mathbf{P}$ and the Gaussian parameters under sparsity regularization~\cite{aksoy2017unmixing}. Letting $\mathcal{N}_\mathbf{x}$ denote the depth-sorted set of Gaussians affecting pixel $\mathbf{x}$ and $\alpha_i'$ each Gaussian's opacity modulated by its 2D spatial footprint, plugging the diffuse decomposition into the 3DGS rendering pipeline~\cite{kerbl20233d} and factoring out $\mathbf{P}$ gives:
\begin{equation}\label{eq:palettegaussian_factored}
\mathbf{C}(\mathbf{x}; \theta, \phi) = \underbrace{\left[\sum_{i \in \mathcal{N}_\mathbf{x}} \mathbf{w}_i^{\top}\, \alpha_i' \prod_{j=1}^{i-1}(1 - \alpha_j') \right] \mathbf{P}}_{\text{diffuse: weighted palette}}  + \underbrace{\sum_{i \in \mathcal{N}_\mathbf{x}} \mathbf{c}_i^s(\theta, \phi)\, \alpha_i' \prod_{j=1}^{i-1}(1 - \alpha_j')}_{\text{specular}}.
\end{equation}

This formulation reveals two fundamental limitations of primitive-space decomposition. \emph{First}, the pixel-level palette weights (first term of Eq.~\ref{eq:palettegaussian_factored}) are alpha-blended per-Gaussian weights $\mathbf{w}_i$. Even when each $\mathbf{w}_i$ is sparse and sums to one, blending many Gaussians at a pixel breaks sparsity, producing nonzero weights across most palette colors and causing edits to bleed into unintended regions (Fig.~\ref{fig:sparsity-compare}, and supplemental). In regions of sparse coverage, the residual $(1-\sum_j w_j)$ contributes black, diluting editing control. \emph{Second}, because $\mathbf{P}$ affects only the diffuse term while specular highlights remain independent, enforcing pixel-level color constraints is ill-posed: optimizing the palette alone causes bleeding; adding regularization still bleeds, converges slowly (over 10 seconds on GPU), and drives the palette to extreme values; jointly optimizing the palette and specular SH converges faster but shows severe bleeds, as SH encode global view-dependent
effects with no clear strategy to impose local regularization (see Fig.~6 in supplemental).

\section{Method: Representation and Finetuning}
\label{sec:method_train}
Our goal is to enable real-time, high-quality palette-based editing of 3DGS scenes through three controls: direct palette manipulation, per-palette tone curves for independent luminance adjustment, and pixel-level color constraints that propagate consistently across novel views (Figs.~\ref{fig:teaser} and~\ref{fig:gallery1}). This requires two design choices. \emph{First}, since edits are specified and perceived at the pixel-level, we splat palette weights rather than colors (\S\ref{sec:repr}) and regularize them in view-space for sparsity and color consistency (\S\ref{sec:barycentric_supervision}). \emph{Second}, to remain compatible with vanilla 3DGS, we reparametrize a pretrained vanilla 3DGS by freezing geometric parameters and replacing the color SH with palette-weight SH, jointly optimized with the palette (\S\ref{sec:optimization}). This keeps the palette linear in the rendering pipeline, enabling closed-form bake-back for palette edits and a brief finetuning step for tone-curve edits. \rev{An overview of our method is shown in Fig.~\ref{fig:pipeline}.} Editing is deferred to \S\ref{sec:method_editing}.

\subsection{View-space Palette Decomposition}
\label{sec:repr}
We work entirely in RGB-space throughout the framework. Given a pretrained vanilla 3DGS, we fix all geometric parameters (position $\boldsymbol{\mu}_i$, covariance $\boldsymbol{\Sigma}_i$, opacity $\alpha_i$) and finetune\footnote{Although the SH coefficients are randomly initialized, we refer to this stage as finetuning because it adapts the pretrained Gaussian geometry to a new appearance representation rather than training a 3DGS from scratch.} only the per-Gaussian SH coefficients to encode view-dependent palette weights $\mathbf{w}_i \in \mathbb{R}^{K}$ instead of RGB colors, where $K$ is the total number of palette colors. The palette $\mathbf{P} = [\vmathbb{0}, \vmathbb{1}, \mathbf{p}_1, \ldots, \mathbf{p}_{K-2}] \in \mathbb{R}^{K \times 3}$ has two fixed achromatic anchors (black $\vmathbb{0}$ and white $\vmathbb{1}$) and $K-2$ learnable chromatic vertices $\mathbf{p}_i \in [0, 1]^3$, so that black and white are always representable without consuming chromatic capacity. For a viewpoint $(\theta, \phi)$, each Gaussian's weight vector is $\mathbf{w}_i(\theta, \phi) = \sum_{l,m} f^{w}_{l,m,i}\, Y_l^m(\theta, \phi) \in \mathbb{R}^{K}$, where $f^{w}_{l,m,i} \in \mathbb{R}^{K}$ are learned SH coefficients. Letting $\beta_i(\mathbf{x}; \theta, \phi) = \alpha_i'\, \prod_{j=1}^{i-1}(1 - \alpha_j')$ denote the alpha-blending coefficient at pixel $\mathbf{x}$ (a fixed constant under our frozen geometry), splatting these vectors yields per-pixel weights
\begin{equation}\label{eq:splatted_weights}
\mathbf{W}(\mathbf{x}; \theta, \phi) = \sum_{i \in \mathcal{N}_\mathbf{x}} \beta_i(\mathbf{x}; \theta, \phi)\, \mathbf{w}_i(\theta, \phi),
\end{equation}
from which the rendered pixel color is recovered as $\mathbf{P}^\top \mathbf{W}(\mathbf{x}; \theta, \phi)$. The combination of frozen $\beta_i$ and a linear SH-to-weights map keeps the rendering pipeline linear in the palette, allowing us to bake any palette edit back into vanilla 3DGS at no additional cost (see supplemental for full proof).

\subsection{Barycentric Supervision}
\label{sec:barycentric_supervision}
The view-space weights of \S\ref{sec:repr} are directly available for regularization, but two further properties are critical for high-quality palette editing: sparsity, which prevents edits from bleeding across colors, and color consistency, which ensures similar colors propagate edits identically (Fig.~6 in supplemental). We achieve both properties simultaneously by supervising the splatted weights $\mathbf{W}(\mathbf{x}; \theta, \phi)$ against the \emph{barycentric decomposition} of the frozen splatted color $\mathbf{c}(\mathbf{x}; \theta, \phi) \in \mathbb{R}^3$ of the vanilla 3DGS over the current palette $\mathbf{P}$: barycentric coordinates are inherently 4-sparse for any color inside a tetrahedral wedge, and similar colors lying within the same wedge receive nearby barycentric coordinates by construction.

To realize this barycentric decomposition, we tessellate $\mathbf{P}$ into $K-2$ tetrahedral wedges $T_i = (\mathbf{0}, \mathbf{1}, \mathbf{p}_i, \mathbf{p}_{i+1})$, $i = 1, \ldots, K-2$ with wraparound on $i+1$, each fanning around the line of greys \cite{tan2018efficient}. Choosing which wedge a given color belongs to requires the chromatic vertices to be ordered around the grey axis, and re-sorting them after every gradient step would break the optimization. We sidestep this by parameterizing each chromatic vertex $\mathbf{p}_i$ in HSL-space with hue $h_i \in [0, 2\pi)$, saturation $s_i \in [0, 1]$, and lightness $l_i \in [0, 1]$, recovering its RGB position via the standard HSL-to-RGB transform. Hues are represented as cumulative offsets from a learnable base angle $h_0$:
$h_i = (h_0 + \sum_{k < i} g_k) \bmod 2\pi$,
where the gaps $g_k = 2\pi \cdot \mathrm{softplus}(\tilde{g}_k) / \sum_j \mathrm{softplus}(\tilde{g}_j)$ are softplus-normalized to sum to $2\pi$, so the $\mathbf{p}_i$ are always in counter-clockwise hue order around the achromatic axis without explicit sorting.

\begin{figure}
    \centering
    \includegraphics[width=\linewidth]{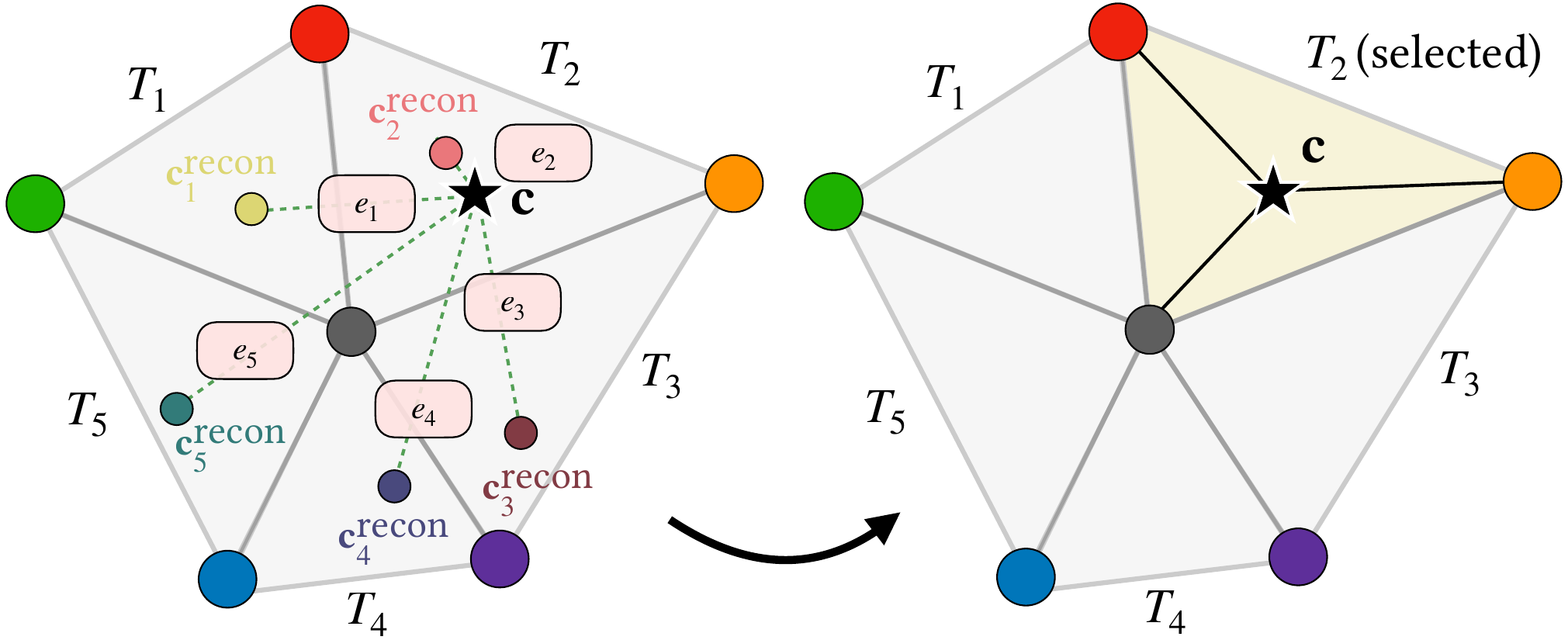}
    \caption{\textbf{Differentiable barycentric soft wedge selection (\S\ref{sec:barycentric_supervision}).} The palette is tessellated into $K-2$ tetrahedral wedges $T_i = (\mathbf{0}, \mathbf{1}, \mathbf{p}_i, \mathbf{p}_{i+1})$ fanning around the grey axis (shown in 2D, with the grey axis collapsed to the central point). \textit{Left:} for each wedge we solve for the raw barycentric coordinates of the query color $\mathbf{c}$ and project them onto the simplex via softplus to obtain a reconstruction $\mathbf{c}_i^{\mathrm{recon}}$; the containing wedge ($T_2$) yields $\mathbf{c}_i^{\mathrm{recon}} \approx \mathbf{c}$, while the others are pulled toward their interior. \textit{Right:} a softmax over $-e_i / \tau$ with $e_i = \|\mathbf{c}_i^{\mathrm{recon}} - \mathbf{c}\|_2^2$ selects the containing wedge crisply, and the target weights $\mathbf{W}_{\mathrm{bary}}$ are formed from its raw barycentric coordinates.}

    \label{fig:bary_selection}
\end{figure}

With the wedge tessellation well-defined throughout optimization, we construct \emph{target barycentric coordinates} $\mathbf{W}_{\mathrm{bary}}(\mathbf{x}; \theta, \phi) \in \mathbb{R}^{K}$ that decompose $\mathbf{c}(\mathbf{x}; \theta, \phi)$ over $\mathbf{P}$. A 2D illustration of the algorithm can be seen in Fig.~\ref{fig:bary_selection}. For each wedge $T_i$, we solve a $3 \times 3$ linear system to obtain the unique barycentric coordinates $(a_i, b_i, c_i, d_i)$ satisfying $\mathbf{c} = a_i \mathbf{0} + b_i \mathbf{1} + c_i \mathbf{p}_i + d_i \mathbf{p}_{i+1}$ and $a_i + b_i + c_i + d_i = 1$. Choosing which wedge $\mathbf{c}$ belongs to is non-trivial because $\mathbf{c}$ may lie outside any single wedge, in which case some of $(a_i, b_i, c_i, d_i)$ are negative. We therefore use the softplus simplex projection only as a tool for \emph{soft wedge selection}: we project each set of coordinates onto the simplex via $\widetilde{a}_i = \mathrm{softplus}(a_i) / (\mathrm{softplus}(a_i) + \mathrm{softplus}(b_i) + \mathrm{softplus}(c_i) + \mathrm{softplus}(d_i))$ and similarly for $\widetilde{b}_i, \widetilde{c}_i, \widetilde{d}_i$, reconstruct each wedge's color as $\mathbf{c}_i^{\mathrm{recon}} = \widetilde{a}_i \mathbf{0} + \widetilde{b}_i \mathbf{1} + \widetilde{c}_i \mathbf{p}_i + \widetilde{d}_i \mathbf{p}_{i+1}$, measure the wedge's reconstruction error $e_i = \|\mathbf{c}_i^{\mathrm{recon}} - \mathbf{c}\|_2^2$, and softly select the best-fitting wedge by softmax over errors, $\pi_i = \exp(-e_i / \tau) / \sum_{j=1}^{K-2} \exp(-e_j / \tau)$,
with temperature $\tau = 10^{-9}$. The final target weights are a soft mixture of the \emph{raw} (unprojected) barycentric coordinates across all wedges, $\mathbf{W}_{\mathrm{bary}}(\mathbf{x}; \theta, \phi) = \sum_{i} \pi_i \mathbf{w}_i^{\mathrm{raw}}$, where $\mathbf{w}_i^{\mathrm{raw}} \in \mathbb{R}^{K}$ places $(a_i, b_i, c_i, d_i)$ into the four entries indexing the vertices of wedge $T_i$ (i.e., the entries for $\mathbf{0}, \mathbf{1}, \mathbf{p}_i, \mathbf{p}_{i+1}$) and zeros elsewhere.
This choice ensures that for any color $\mathbf{c}$ strictly inside a wedge $T_{i^*}$, the targets recover $\mathbf{c}$'s true barycentric coordinates exactly in the small-$\tau$ limit. The full computation is differentiable in $\mathbf{P}$, so gradients flow back to the chromatic vertices during finetuning.

\subsection{Joint Palette and Weight SH Finetuning}
\label{sec:optimization}
We finetune the per-Gaussian SH coefficients $\{f^{w}_{l,m,i}\}$ and palette parameters $\{h_0, \tilde{g}_k, s_i, l_i\}$ jointly against the barycentric targets, minimizing
$\mathcal{L} = \mathcal{L}_{\mathrm{color}} + \lambda_{\mathrm{pos}}\,\mathcal{L}_{\mathrm{pos}} + \lambda_{\mathrm{compact}}\,\mathcal{L}_{\mathrm{compact}} + \lambda_{\mathrm{sep}}\,\mathcal{L}_{\mathrm{sep}}$,
where $\mathcal{L}_{\mathrm{color}} = \frac{1}{HW} \sum_{\mathbf{x}} \|\mathbf{W}(\mathbf{x}; \theta, \phi) - \mathbf{W}_{\mathrm{bary}}(\mathbf{x}; \theta, \phi)\|_2^2$ is the barycentric supervision loss (\S~\ref{sec:barycentric_supervision}).
%The auxiliary terms regularize the splatted weights and palette geometry: $\mathcal{L}_{\mathrm{pos}} = \mathrm{mean}(\mathrm{ReLU}(-\mathbf{W}))$ discourages negative splatted weights without breaking the linearity of the SH-to-weights map required for closed-form bake-back; $\mathcal{L}_{\mathrm{compact}} = \sum_{i=1}^{K-2} \mathrm{vol}(T_i)$ sums wedge volumes to counterbalance $\mathcal{L}_{\mathrm{color}}$'s tendency to produce non-representative palettes; and $\mathcal{L}_{\mathrm{sep}}$ penalizes any hue gap $g_k$ below a minimum fraction of the even spacing $2\pi/(K-2)$, preventing palette collapse. We ramp up $\lambda_{\mathrm{compact}}$ late in training so the palette first expands toward the scene's color distribution before being tightened. Hyperparameter values, the schedule, and additional implementation details are in \S\ref{sec:results}.
The auxiliary terms regularize the splatted weights and the palette geometry: $\mathcal{L}_{\mathrm{pos}} = \mathrm{mean}(\mathrm{ReLU}(-\mathbf{W}))$ discourages negative splatted weights without breaking the linearity of the SH-to-weights map required for closed-form bake-back; $\mathcal{L}_{\mathrm{compact}} = \sum_{i=1}^{K-2} \mathrm{vol}(T_i)$ is the sum of wedge volumes, counterbalancing $\mathcal{L}_{\mathrm{color}}$'s tendency to create non-representative palette; and $\mathcal{L}_{\mathrm{sep}}$ penalizes any hue gap $g_k$ falling below a minimum fraction of the even spacing $2\pi/(K-2)$, preventing palette from collapsing. We schedule $\lambda_{\mathrm{compact}}$ so that compactness pressure ramps up in the late stage of training, allowing the palette to first expand toward the scene's color distribution before being tightened. Full hyperparameter values, the schedule, and additional implementation details are provided in \S\ref{sec:results}.

\subsection{Rank-1 Factorization of Higher-Order Weight SH}
\label{sec:low_rank}
\begin{figure}[t]
\includegraphics[width=\linewidth,scale=1]{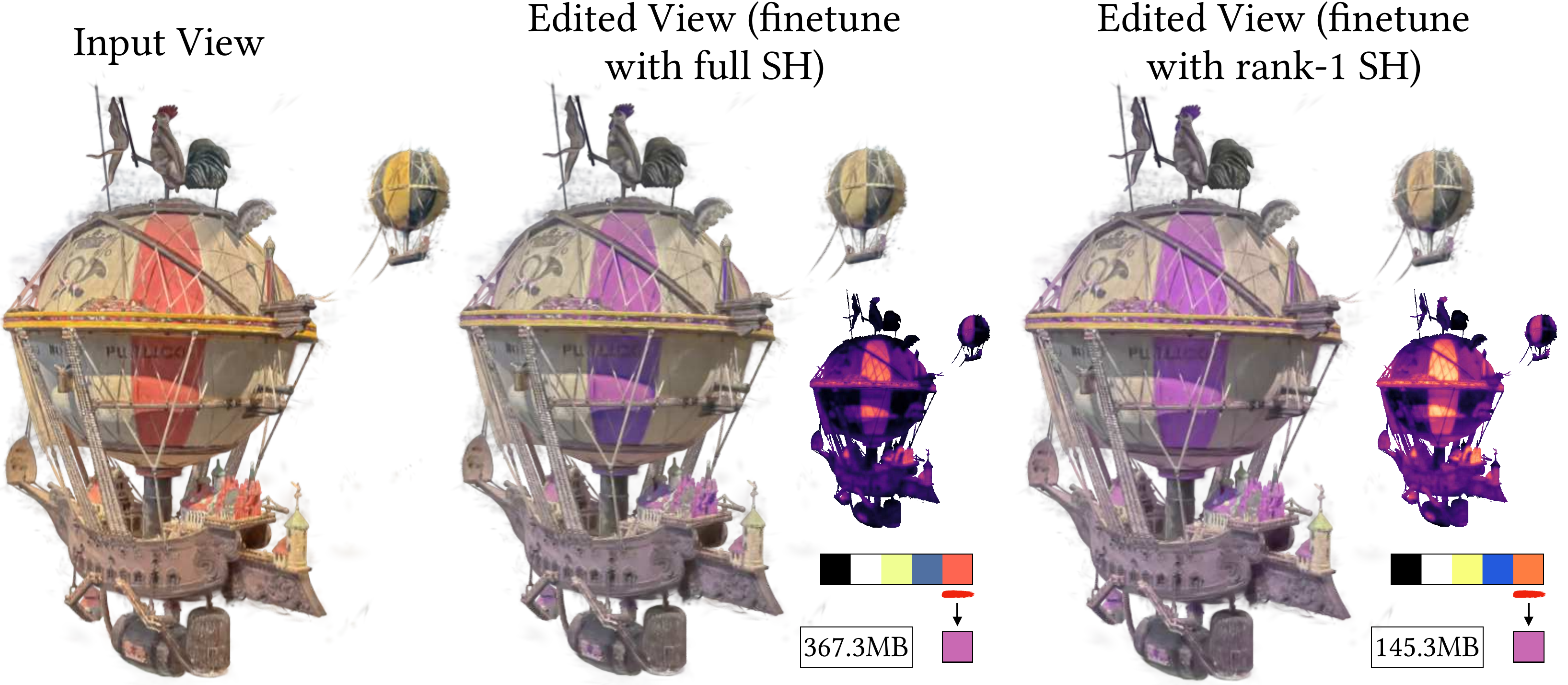}
\caption{
    \textbf{Full vs.\ rank-1 SH for recoloring.} The full SH parametrization (\emph{middle}) and rank-1 factorization (\emph{right}) recover nearly identical palettes; editing the same palette slot from orange to purple (red underline) produces visually comparable results under both, differing only in fine detail. \rev{The FLIP difference maps \cite{andersson2020flip}, each comparing a method's edit to its own unedited rendering, highlight the same regions under full SH and rank-1, with only slightly higher intensity for rank-1; this confirms that the recolor localizes consistently across both.} Storage for each model is annotated beneath.
    \emph{Scene:} \href{https://superspl.at/scene/bf754dec}{``Airship''} by \href{https://superspl.at/user?id=austinbeaulier}{Austin Beaulier}.
}
\label{fig:recolor-diff-rank}
\end{figure}

\begin{figure}[t]
\includegraphics[width=\linewidth,scale=1]{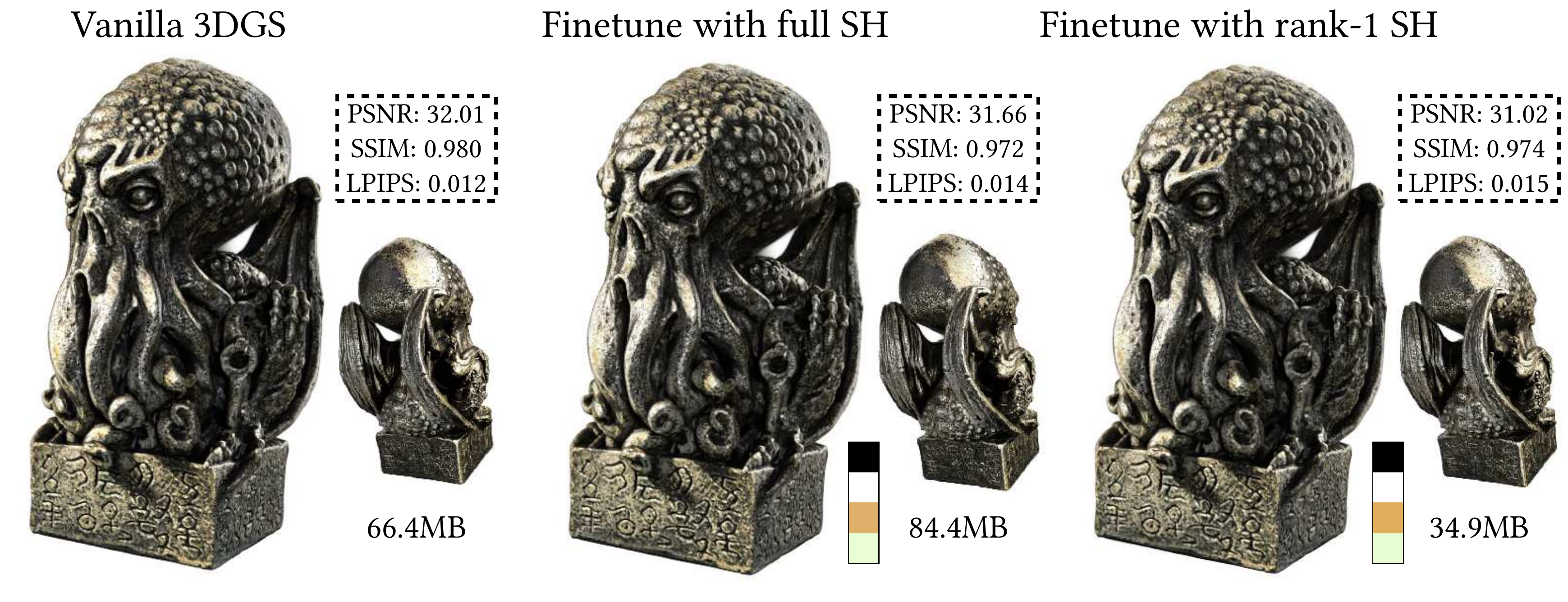}
\caption{
    \rev{
    Rank-1 SH on highly reflective content
    preserves specular highlights at much lower memory. On the metallic \emph{Cthulhu Statue}, our rank-1 factorization (\emph{right}) costs little reconstruction quality relative to full SH (\emph{middle}, ${\sim}0.6$~dB PSNR) while using under half the storage and less than vanilla 3DGS (\emph{left}).
    Palette swatches and per-model storage are annotated.
    \emph{Scene:} \href{https://superspl.at/scene/44e8fe8f}{``Cthulhu Statue''} by \href{https://superspl.at/user/schindelar3d}{Christoph Schindelar}.
    }
}
\label{fig:reflective-rank1}
\end{figure}

The representation in \S\ref{sec:repr} replaces each Gaussian's color SH with $K$-channel weight SH, blowing up per-Gaussian SH storage by a factor of $K/3$ relative to vanilla 3DGS ($2\times$ for $K=6$). To mitigate this, we factorize the higher-order ($l \geq 1$) weight SH coefficients via a reshaped outer product, leaving the band-zero ($l=0$) term unfactorized. Empirically, this gives negligible reconstruction loss compared to finetuning with the full weight SH (Table~\ref{tab:nvs_comparison}) and produces visually comparable recoloring results, though not pixel-identical (Fig.~\ref{fig:recolor-diff-rank}). \rev{This holds even on highly reflective, view-dependent content, where the factorization preserves specular highlights with little quality loss at lower memory (Fig.~\ref{fig:reflective-rank1}).} To describe the factorization, we first fix notation for the unfactorized form. Let $L$ denote the maximum SH band; each Gaussian $i$ has $(L+1)^2 - 1$ higher-order SH coefficients per channel, and stacking all $K$ channels gives a matrix $\mathbf{M}_i \in \mathbb{R}^{K \times ((L+1)^2 - 1)}$. The view-dependent weight vector at direction $(\theta, \phi)$ is
\begin{equation}\label{eq:full_rank_render}
\mathbf{w}_i(\theta, \phi) = \mathbf{f}^{w,0}_i + \mathbf{M}_i\, \mathbf{Y}^{\geq 1}(\theta, \phi),
\end{equation}
where $\mathbf{f}^{w,0}_i \in \mathbb{R}^{K}$ holds the band-zero coefficients and $\mathbf{Y}^{\geq 1}(\theta, \phi) \in \mathbb{R}^{(L+1)^2 - 1}$ are the higher-order SH basis values. To factorize, choose positive integers $P, Q \geq 1$ with $P \cdot Q = K \cdot ((L+1)^2 - 1)$ and store $\mathbf{a}_i \in \mathbb{R}^{P}$, $\mathbf{b}_i \in \mathbb{R}^{Q}$ per Gaussian. We form $\mathbf{S}_i = \mathrm{vec}(\mathbf{a}_i \mathbf{b}_i^\top) \in \mathbb{R}^{PQ}$ via row-major vectorization, and reshape it into the approximated coefficient matrix $\widehat{\mathbf{M}}_i \in \mathbb{R}^{K \times ((L+1)^2 - 1)}$ by
\begin{equation}\label{eq:lowrank_recon}
\widehat{\mathbf{M}}_i[k, lm] = \mathbf{S}_i[lm \cdot K + k],
\end{equation}
for $0 \leq k < K$ and $0 \leq lm < (L+1)^2 - 1$, so successive $K$-chunks of $\mathbf{S}_i$ fill the columns of $\widehat{\mathbf{M}}_i$. The approximated weight vector follows Eq.~\ref{eq:full_rank_render} with $\mathbf{M}_i$ replaced by $\widehat{\mathbf{M}}_i$. Note that $\mathbf{a}_i \mathbf{b}_i^\top$ is rank-1 but $\widehat{\mathbf{M}}_i$ in general is not.
% \yotam{Is it still rank-1 after the reshape? What is the motivation for the reshape?} \ted{no, it's not rank-1 after reshape in general.} 
This reduces per-Gaussian SH storage from $K \cdot (L+1)^2$ to $K + P + Q$. By AM-GM, $P + Q$ is minimized at $P = Q$ subject to $P \cdot Q = K \cdot ((L+1)^2 - 1)$. For $K=6$, $L=3$, the closest integer factorization is $(P, Q) = (9, 10)$, giving $25$ scalars per Gaussian, roughly half of vanilla 3DGS's $48$ (Table~\ref{tab:nvs_comparison}).

\section{Method: Editing}
\label{sec:method_editing}
With the palette and per-Gaussian weight SH finetuned (\S\ref{sec:method_train}), users edit the scene in real-time via three constraint types: pixel-level color constraints (clicking a point in any rendered view and choosing a target color), palette constraints (modifying a palette), or per-palette tone-curve constraints (placing control points on a curve). All three feed a single optimization that seeks the smallest change to the palette and tone curves satisfying all constraints simultaneously, leaving the splatted weights and Gaussian parameters fixed. We define the editing model and constraint types (\S\ref{sec:editing_model}), then describe a constraint-driven solver based on iteratively reweighted least squares (IRLS) and damped block coordinate descent (dBCD) (\S\ref{sec:editing_solver}).

\subsection{Editing Model and Constraints}
\label{sec:editing_model}

Let $\mathbf{P} \in \mathbb{R}^{K \times 3}$ denote the trained palette and $\mathbf{W}(\mathbf{x}; \theta, \phi) \in \mathbb{R}^{K}$ the splatted weights at pixel $\mathbf{x}$ from viewpoint $(\theta, \phi)$. We associate each palette vertex $\mathbf{p}_i$ with a tone curve $f_i: [0, 1] \to [0, 1]$ defined as a biharmonic function~\cite{chao2023colorfulcurves} parametrized by user-placed control points; initially every $f_i$ is an identity function and the palette is unchanged. We discretize each $f_i$ at $S$ equally spaced sample points and let $\mathbf{L}_i \in \mathbb{R}^{S}$ denote the vector of sampled values. The free editing variables are the palette modification $\Delta \mathbf{P} \in \mathbb{R}^{K \times 3}$ and the tone-curve sample vectors $\{\mathbf{L}_i\}_{i=1}^{K}$, with all other parameters of the trained representation held fixed.

These two sets of variables determine the edited color at every pixel. We realize luminance editing by redistributing weight mass between the black ($\vmathbb{0}$) and white ($\vmathbb{1}$) palette of $\mathbf{W}$. Concretely, given a viewpoint $(\theta, \phi)$ and a pixel $\mathbf{x}$, the splatted weights are $\mathbf{W}(\mathbf{x}; \theta, \phi) = (W_1, W_2, W_3, \ldots, W_K)$ with index $1$ for black and index $2$ for white, and the edited weights are (dropping the $(\mathbf{x}; \theta, \phi)$ for compactness)
\begin{equation}\label{eq:edit_weight_shift}
W_1' = W_1 - \delta, \qquad W_2' = W_2 + \delta, \qquad W_i' = W_i \;\;\text{for}\;\; i \geq 3,
\end{equation}
where the luminance shift $\delta(\mathbf{x}; \theta, \phi)$ is computed from the per-pixel luminance $L_0(\mathbf{x}; \theta, \phi) = \mathbf{w}_L^\top \mathbf{C}_0(\mathbf{x}; \theta, \phi)$ (BT.709~\cite{itu_bt709}, $\mathbf{w}_L = (0.2126, 0.7152, 0.0722)^\top$, with $\mathbf{C}_0$ the original rendered color) and the tone-curve samples as
\begin{equation}\label{eq:edit_delta}
\delta(\mathbf{x}; \theta, \phi) = \sum_{i=1}^{K} \widetilde{W}_i(\mathbf{x}; \theta, \phi) \cdot \big( f_i(L_0(\mathbf{x}; \theta, \phi)) - L_0(\mathbf{x}; \theta, \phi) \big),
\end{equation}
Here $\widetilde{W}_i$ is obtained by scaling the black and white entries of $\mathbf{W}$ by a small factor (we use $\eta = 0.01$) and renormalizing the result to sum to one, so chromatic palette colors dominate the shift. The edited color then follows from the modified weights and the edited palette is,
\begin{equation}\label{eq:edit_color}
\mathbf{C}'(\mathbf{x}; \theta, \phi) = (\mathbf{P} + \Delta \mathbf{P})^\top \mathbf{W}'(\mathbf{x}; \theta, \phi).
\end{equation}
We prove in the supplemental materials that the weight shift alone (Eq.~\ref{eq:edit_weight_shift}, with $\Delta \mathbf{P} = \mathbf{0}$) is equivalent to a per-pixel luminance edit applied in proportion to each palette color's chromatic weight: the resulting per-pixel BT.709 luminance is the chromatic-weight-blended target $L_0'(\mathbf{x}) = \sum_{i \geq 3} \widetilde{W}_i\, f_i(L_0(\mathbf{x}))$, and the chromatic component of the rendered color is preserved exactly. When black and white are kept fixed ($\Delta\mathbf{P}_1 = \Delta\mathbf{P}_2 = \mathbf{0}$), Eq.~\ref{eq:edit_color} simplifies to
\begin{equation}\label{eq:edit_color_grey}
\mathbf{C}'(\mathbf{x}; \theta, \phi) = (\mathbf{P} + \Delta \mathbf{P})^\top \mathbf{W}(\mathbf{x}; \theta, \phi) + \delta(\mathbf{x}; \theta, \phi) % \cdot \mathbf{1}
\vmathbb{1}
\end{equation}

% \yotam{Is $\delta$ a scalar? Equation \ref{eq:edit_weight_shift} suggests that it is. But then why is it multiplied by 1?}  \ted{It is scalar. The 1 is the white vector defined above.}

Given this editing model, users specify the desired edit via three types of constraints. A \emph{pixel-level color constraint} is a tuple $(\mathbf{x}_n,\allowbreak (\theta_n, \phi_n),\allowbreak \mathbf{c}_n^{\mathrm{tgt}})$ specifying that the edited color $\mathbf{C}'(\mathbf{x}_n; \theta_n, \phi_n)$, evaluated using the splatted weights $\mathbf{W}(\mathbf{x}_n; \theta_n, \phi_n)$ at that pixel $\mathbf{x}_n$, should equal a target color $\mathbf{c}_n^{\mathrm{tgt}}$. A \emph{palette constraint} is a pair $(i, \mathbf{p}_i^{\mathrm{tgt}})$ specifying that the edited vertex $\mathbf{p}_i + \Delta \mathbf{p}_i$ should equal $\mathbf{p}_i^{\mathrm{tgt}}$.
% \yotam{Can we really change the black and white anchors if they are kept fixed during the luminance shift solve?} \ted{Yes they can, but should we say it?}
A \emph{tone-curve constraint} is a tuple $(i, x_n, y_n)$ specifying that $f_i(x_n) = y_n$, i.e., the user has placed a control point at $(x_n, y_n)$ on the tone curve of palette color $i$.

\subsection{Constraint-Driven Solver}
\label{sec:editing_solver}
Given $N_{\mathrm{img}}$ pixel-level color constraints, $N_p$ palette constraints, and $N_l$ tone-curve constraints, our goal is to find the smallest change to $\Delta \mathbf{P} \in \mathbb{R}^{K \times 3}$ and  $\{\mathbf{L}_i \in \mathbb{R}^{S} \}_{i=1}^{K}$ that satisfies all of them. We use $\mathbf{B} \in \mathbb{R}^{S \times S}$ to denote the discrete Laplacian with mirroring at endpoints for natural boundary conditions. The editing problem is a sparsity-regularized constrained optimization\footnote{We use $w_{\mathrm{eq}} = 10^3$ to enforce constraints stiffly and $w_{\mathrm{sp}} = 0.1$.}:
\begin{equation}\label{eq:edit_full_energy}
\min_{\{\mathbf{L}_i\},\,\Delta \mathbf{P}} \quad E_{\mathrm{sp}} \;+\; w_{\mathrm{eq}}\,\big(E_{\mathrm{img}} + E_{l} + E_{p}\big),
\end{equation}
subject to $\mathbf{L}_{i,1} = 0$ and $\mathbf{L}_{i,S} = 1$ for all $i$ (pinning the first and last samples of each tone curve to $0$ and $1$), where $E_{\mathrm{sp}}$ is the $L_{2,1}$ sparsity term \cite{chao2023colorfulcurves}:
\begin{equation}\label{eq:edit_sparsity}
E_{\mathrm{sp}} \;=\; \sum_{i=1}^{K} \sqrt{\, \mathbf{L}_i^\top \mathbf{B}^\top \mathbf{B}\, \mathbf{L}_i \;+\; w_{\mathrm{sp}} \cdot \|\Delta \mathbf{p}_i\|_2^2\,},
\end{equation}
which induces global sparsity across palette colors and per-palette tone curves. The pixel-level color constraint loss $E_{\mathrm{img}}$ aggregates the squared color residual at every constraint pixel,
\begin{equation}\label{eq:edit_Eimg}
E_{\mathrm{img}} \;=\; \sum_{n=1}^{N_{\mathrm{img}}} \big\| \mathbf{C}'(\mathbf{x}_n; \theta_n, \phi_n) - \mathbf{c}_n^{\mathrm{tgt}} \big\|_2^2,
\end{equation}
with $\mathbf{C}'$ from Eq.~\ref{eq:edit_color}. The tone-curve and palette constraint losses are the squared discrepancies against their targets,
\begin{equation}\label{eq:edit_El_Ep}
E_l \;=\; \sum_{i=1}^{K} \|\mathbf{S}_i \odot \mathbf{L}_i - \mathbf{L}_i^{\mathrm{tgt}}\|_2^2, \qquad E_p \;=\; \sum_{i \in P_c} \big\| (\mathbf{p}_{i} + \Delta \mathbf{p}_{i}) - \mathbf{p}_{i}^{\mathrm{tgt}} \big\|_2^2,
\end{equation}
where $\mathbf{S}_i, \mathbf{L}_i^{\mathrm{tgt}} \in \mathbb{R}^S$ are selection/target vectors over curve $i$'s sample positions, and $P_c$ is the index set of constrained palette colors.
User constraints on black and white ($i \in \{1, 2\}$) bypass $E_p$ and directly update $\mathbf{P}$,
% Constraints on black and white ($i \in \{1, 2\}$) bypass $E_p$ and are pinned directly to any user-specified targets at each iteration,
preserving the weight-shift equivalence (Eq.~\ref{eq:edit_color_grey}).

Our $E_{\mathrm{img}}$ couples the luminance and palette blocks in RGB-space: both tone curve contributions (via $\delta(\mathbf{x}) \cdot \vmathbb{1}$ in Eq.~\ref{eq:edit_color_grey}) and palette shifts (via $\sum_i W_i(\mathbf{x}) \Delta \mathbf{p}_i$) contribute to the same color residual, and chromatic palette edits can produce nonzero projections onto the grey axis. \citet{chao2023colorfulcurves} avoid this coupling by working in Lab-space, where tone curves affect only $L^*$ and palette shifts only $ab$, so the $E_{\mathrm{img}}$ residual separates along these axes. This orthogonality lets them apply BCD on lightness and chroma as independent sub-problems, yielding a fast solver. We thoroughly investigated a Lab-space formulation (supplemental), but adopting it requires a non-linear Lab-to-RGB map at edit time and breaks closed-form bake-back for palette edits. We therefore stay in RGB-space and design a tailored solver that exploits two structural properties of the problem.

%%%%% Ted revise and proofread above all
A direct general-purpose solver would run far too slowly for interaction, because of the problem size: Eq.~\ref{eq:edit_full_energy} has $SK + 3(K-2)$ unknowns (e.g., ${\sim} 600$ for $S = 100$, $K = 6$).
The first property comes from the $L_{2,1}$ square root in $E_{\mathrm{sp}}$ (Eq.~\ref{eq:edit_sparsity}). Let $a_i = \mathbf{L}_i^\top \mathbf{B}^\top \mathbf{B}\, \mathbf{L}_i + w_{\mathrm{sp}} \|\Delta \mathbf{p}_i\|_2^2$ denote the argument of the $i$-th square root, so $E_{\mathrm{sp}} = \sum_i \sqrt{a_i}$. The square root is non-differentiable at $a_i = 0$ and produces a $\sqrt{a_i}$-denominator in the gradient, making the sub-problem in $(\mathbf{L}, \Delta\mathbf{P})$ non-linear. We sidestep both issues via the variational identity
% \yot{This should be $\sqrt{a_i} = d_i^* = \textrm{argmin} \dots$, right? Maybe clause after this equation defining $d_i*$ isn't necessary?} \ted{I think writing it with min is more accurate.}
\begin{equation}\label{eq:variational}
\sqrt{a_i} \;=\; \min_{d_i > 0}\, \frac{a_i}{2 d_i} + \frac{d_i}{2}, \: \text{which has unique minimizer } d_i^* = \sqrt{a_i}.
\end{equation}
% To verify, set the derivative with respect to $d_i$ to zero: $-a_i / (2 d_i^2) + 1/2 = 0$ gives $d_i = \sqrt{a_i}$, and substituting back yields $\sqrt{a_i}/2 + \sqrt{a_i}/2 = \sqrt{a_i}$.
Introducing auxiliary scalars $\{d_i\}_{i=1}^{K}$ and replacing $E_{\mathrm{sp}}$ with the augmented energy $\widetilde{E}_{\mathrm{sp}}(\mathbf{L}, \Delta\mathbf{P}, \{d_i\}) = \sum_i \big[\tfrac{a_i}{2 d_i} + \tfrac{d_i}{2}\big]$ preserves the minimum but makes the $(\mathbf{L}, \Delta\mathbf{P})$ block quadratic when $\{d_i\}$ are frozen. IRLS alternates between solving for $(\mathbf{L}, \Delta\mathbf{P})$ via a linear system with $\{d_i\}$ frozen and updating each $d_i \leftarrow \sqrt{a_i}$ in closed-form ($\{1/d_i\}$ serve as the IRLS weights). The second property targets the coupling: $\mathbf{C}'(\mathbf{x})$ is affine in $(\mathbf{L}, \Delta\mathbf{P})$ jointly (Eq.~\ref{eq:edit_color_grey}), so $E_{\mathrm{img}}$ becomes quadratic in each block when the other is held fixed. BCD alternates between an $\mathbf{L}$-step and a $\Delta \mathbf{P}$-step, each treating the other block as constant. Concretely, the pixel constraint $\mathbf{C}'(\mathbf{x}_n) = \mathbf{c}_n^{\mathrm{tgt}}$ (Eq.~\ref{eq:edit_color_grey}) rearranges to $\big(\mathbf{c}_n^{\mathrm{tgt}} - \mathbf{C}_0(\mathbf{x}_n)\big) = \sum_{i=1}^{K} W_i(\mathbf{x}_n) \Delta\mathbf{p}_i + \delta(\mathbf{x}_n) \cdot \vmathbb{1}$, splitting the color residual between palette and luminance contributions. The $\mathbf{L}$-step absorbs the luminance residual that $\Delta \mathbf{P}$ has not covered (projecting onto $\mathbf{w}_L$~\cite{itu_bt709}),
\begin{equation}\label{eq:L_target}
\Delta L^*(\mathbf{x}_n) \;=\; \mathbf{w}_L^\top \big(\mathbf{c}_n^{\mathrm{tgt}} - \mathbf{C}_0(\mathbf{x}_n)\big) \;-\; \mathbf{w}_L^\top \!\!\sum_{i=1}^{K} W_i(\mathbf{x}_n) \Delta \mathbf{p}_i,
\end{equation}
while the $\Delta \mathbf{P}$-step absorbs the full color residual that $\delta$ has not covered,
\begin{equation}\label{eq:P_target}
\mathbf{r}^P(\mathbf{x}_n) \;=\; \big(\mathbf{c}_n^{\mathrm{tgt}} - \mathbf{C}_0(\mathbf{x}_n)\big) \;-\; \delta(\mathbf{x}_n) \cdot \vmathbb{1}.
\end{equation}

The asymmetry is intentional: $\delta$ acts only along the grey axis, so projecting onto luminance (Eq.~\ref{eq:L_target}) determines its target; $\Delta \mathbf{P}$ acts in 3D, so it must see the full residual (Eq.~\ref{eq:P_target}) including the grey-axis component, otherwise the chromatic palette shifts cannot absorb luminance work and $\delta$ alone bears it. This distinguishes our setup from \citet{chao2023colorfulcurves}, whose Lab-space residual decomposes orthogonally so that each block sees an independent projection. Combining IRLS with BCD yields three cheap sub-problems: an $\mathbf{L}$-step that solves a sparse linear system in $SK$ unknowns, a $\Delta \mathbf{P}$-step that solves a small unconstrained quadratic in $3(K-2)$ unknowns via inner IRLS, and a closed-form $d$-step from the variational identity. Plain alternation oscillates because both blocks satisfy the grey-axis component of the image-space residual: tone curves contribute $\delta(\mathbf{x}) \cdot \vmathbb{1}$, chromatic palette shifts contribute the grey-axis projection of $\sum_i W_i(\mathbf{x}) \Delta \mathbf{p}_i$, and each block's update absorbs whatever the other just provided. We damp each block update with a convex combination of the previous iterate, breaking the oscillation and yielding monotone convergence in practice. Convergence typically requires 5--15 outer iterations and tens of milliseconds per edit; explicit forms of the $\mathbf{L}$ and $\Delta \mathbf{P}$ systems, the damping rule, and the full algorithm are provided in the supplemental.

\begin{table*}[t]
\caption{
    Quantitative comparison with vanilla 3DGS~\cite{kerbl20233d} on LLFF~\cite{mildenhall2019local}, Tanks\&Temples~\cite{knapitsch2017tanks}, and NeRF Synthetic~\cite{mildenhall2021nerf}. Our rank-1 factorization (\emph{Ours-R-NC}, \emph{Ours-R-C}) matches the full weight SH (\emph{Ours-Full-NC}) at ${\sim}60\%$ of vanilla 3DGS storage. $K$-channel weight splatting modestly reduces FPS, in exchange for editable view-space palette weights. \emph{NC} (non-clamped) and \emph{C} (clamped) refer to the simplex-clamped barycentric targets in the final finetuning stage (\S\ref{sec:implementation}); \emph{Ours-R-C} trades ${\sim}1$~dB PSNR for non-negative, edit-friendly weights.
}
\label{tab:nvs_comparison}
\centering
\footnotesize
\setlength{\tabcolsep}{3.5pt}
\begin{tabular}{l|cccccc|cccccc|cccccc}
\toprule
Dataset & \multicolumn{6}{c|}{LLFF} & \multicolumn{6}{c|}{Tanks\&Temples} & \multicolumn{6}{c}{NeRF Synthetic} \\
Method$|$Metric & \textit{LPIPS}$\downarrow$ & \textit{SSIM}$\uparrow$ & \textit{PSNR}$\uparrow$ & Train$^\dagger$ & FPS$^\dagger$ & Mem & \textit{LPIPS}$\downarrow$ & \textit{SSIM}$\uparrow$ & \textit{PSNR}$\uparrow$ & Train$^\dagger$ & FPS$^\dagger$ & Mem & \textit{LPIPS}$\downarrow$ & \textit{SSIM}$\uparrow$ & \textit{PSNR}$\uparrow$ &  Train$^\dagger$ & FPS$^\dagger$ & Mem \\
\midrule
Vanilla 3DGS & 
0.207 & 0.782 & 22.84 & 3m22s & 192 & 86.5MB & 
0.146 & 0.828 & 22.05 & 46m37s & 168 & 229MB &
0.033 & 0.960 & 30.50 & 8m33s & 218 & 35.5MB \\
Ours-Full-NC & 
0.220 & 0.769 & 22.34 & 41s$^*$ & 164 & 142MB & 
0.1662 & 0.807 & 21.19 & 2m50s$^*$ & 143 & 417MB & 
0.041 & 0.952 & 30.15 & 56.8s$^*$ & 201 & 54.8MB \\
Ours-R-NC  & 
0.221 & 0.768 & 22.09 & 38.9s$^*$ & 166 & 53.5MB & 
0.169 & 0.797 & 21.42 & 2m28s$^*$ & 143 & 140MB & 
0.044 & 0.948 & 29.61 & 53.5s$^*$ & 202 & 21.7MB  \\
Ours-R-C & 
0.227 & 0.766 & 21.97 & 38.8s$^*$ & 166 & 53.5MB & 
0.174 & 0.792 & 21.12 & 2m28s$^*$ & 143 & 140MB & 
0.047 & 0.945 & 29.03 & 54.9s$^*$ & 201 & 21.7MB  \\
\bottomrule
\addlinespace[2pt]
\multicolumn{19}{@{}l@{}}{\scriptsize $^\dagger$Training on NVIDIA H200; FPS on NVIDIA A100. $^*$``Train'' is finetuning duration only (on pretrained vanilla 3DGS).} \\
\multicolumn{19}{@{}l@{}}{\scriptsize $K = 5$ for all datasets except Tanks\&Temples ($K = 6$). We train vanilla 3DGS with uniform default settings across all scenes.} \\
\end{tabular}
\end{table*}

%%%%%%%%%%%%% Results %%%%%%%%%%%%%

\section{Results and Evaluation}
\label{sec:results}
\begin{figure}[t]
\includegraphics[width=\linewidth]{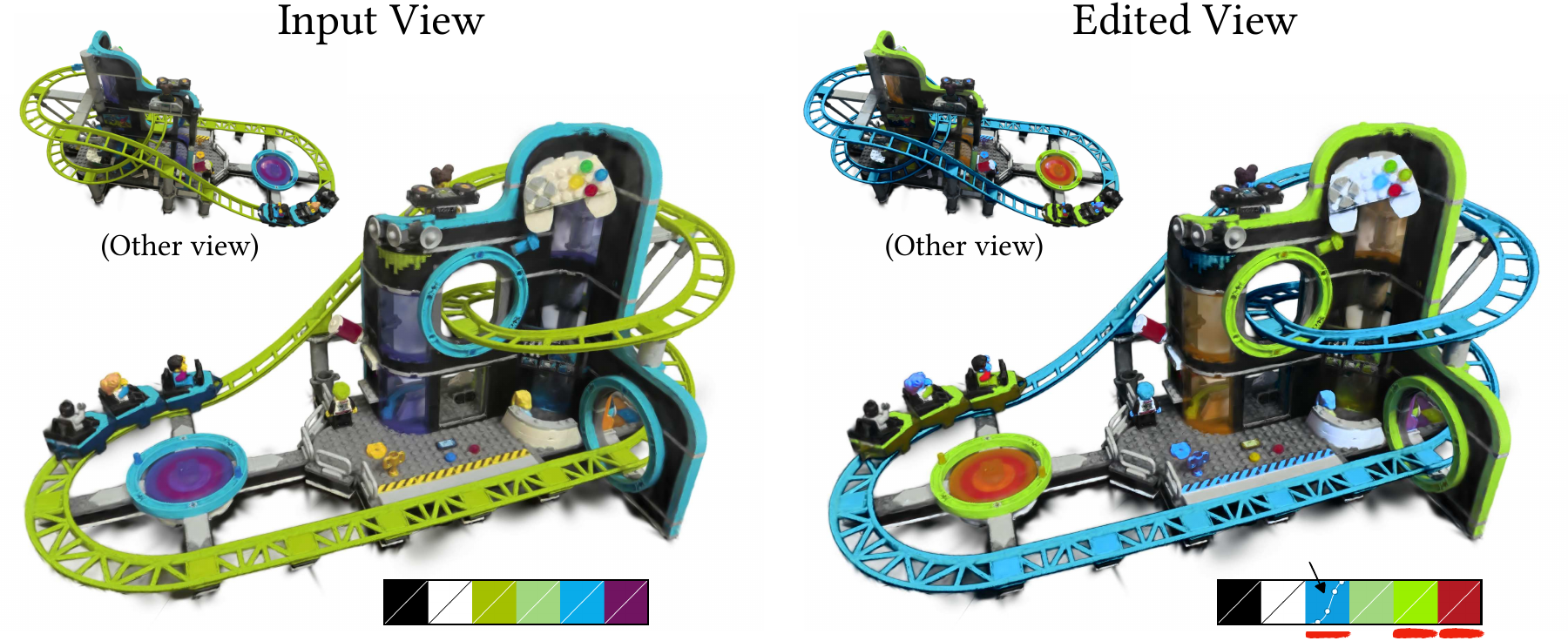}
\caption{
    \rev{
    %Our method provides a sparse palette decomposition of the scene, so edits to a palette color affect only the corresponding regions and remain consistent across views (\emph{other view, inset}).
    Our sparse palette decomposition keeps edits local and view-consistent (\emph{inset}).
    Here we recolor the green track to blue, the decorative frame to green, and the central disc from purple to red by editing three palette entries (red underlines), and additionally adjust the track's contrast with its tone curve. The palette is shown beneath each view.
    \emph{Scene:} \href{https://superspl.at/scene/ca2e3228}{``LEGO roller coaster''} by \href{https://superspl.at/user/wanyanyan}{wanyanyan}.
    }
}
\label{fig:rollercoaster}
\end{figure}

\begin{figure*}[t!]
\includegraphics[width=0.96\linewidth,scale=1]{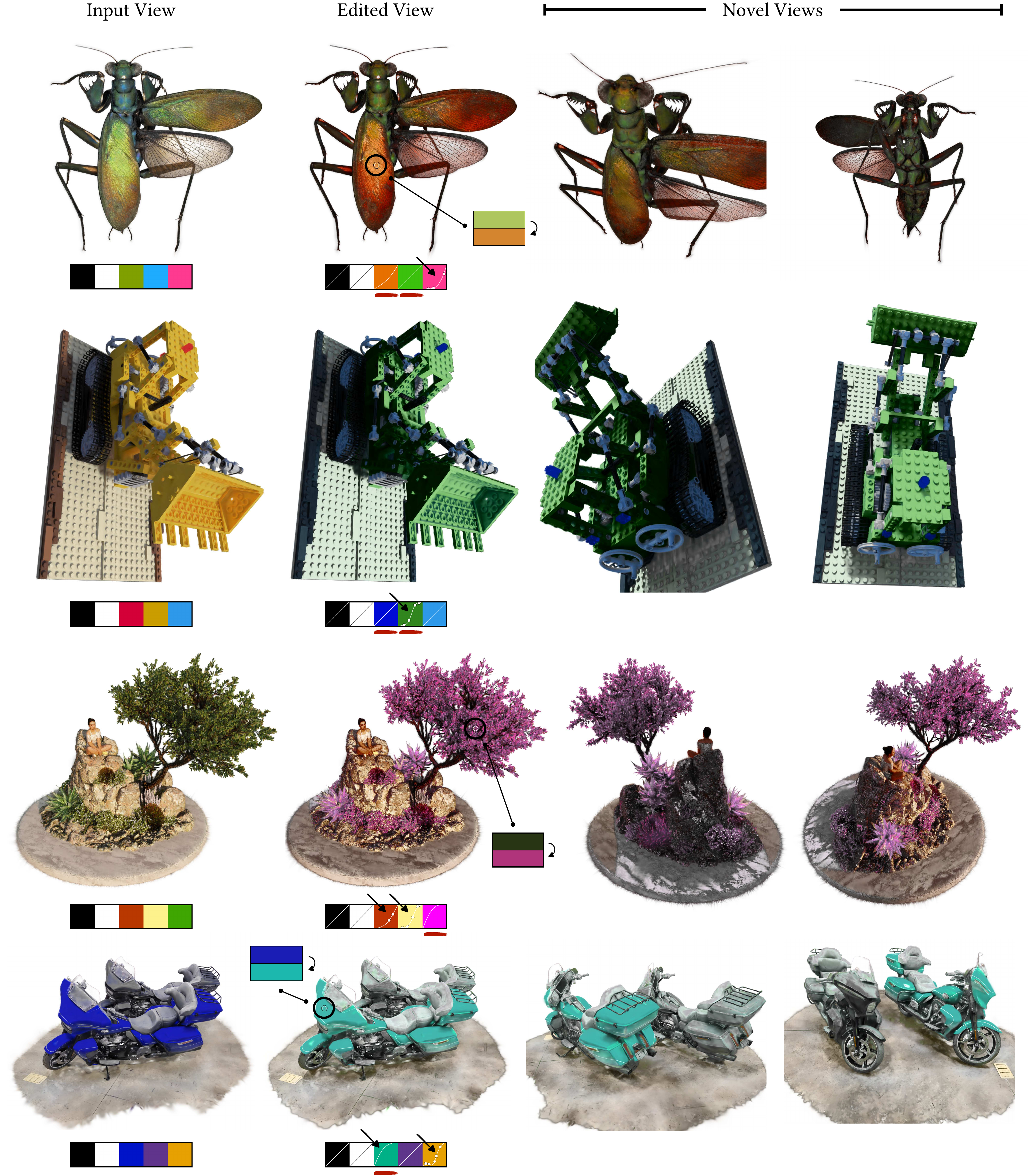}
\caption{\textbf{Gallery 1 of edits.} We edit a variety of scenes through palette modifications (red underlines), per-palette tone curves, and pixel-level color constraints (black circles with specified target colors). Each edited view combines multiple types of edits applied jointly, and propagates consistently across novel views. Additional editing examples can be found in the supplemental materials.
\emph{Scenes (top to bottom):} \href{https://superspl.at/scene/c248dde4}{"Metallyticus splendidus"} by \href{https://superspl.at/user?id=scant3d}{Fabian Plum}, Lego from NeRF synthetic dataset, \href{https://superspl.at/scene/6bbe46da}{"ISLAND – DIORAMA"} by \href{https://superspl.at/user?id=sa3d}{Stéphane Agullo}, \href{https://superspl.at/scene/72d0dc3a}{"Harley-Davidson Street Glide (XGRIDS PortalCam)"} by \href{https://superspl.at/user?id=tosolini}{Paolo Tosolini}.
}
\label{fig:gallery1}
\end{figure*}

We demonstrate a range of editing capabilities: palette-based recoloring (Figs.~\ref{fig:teaser}, \rev{\ref{fig:pipeline}}, \ref{fig:recolor-diff-rank}, \rev{\ref{fig:rollercoaster}}, \ref{fig:gallery1}), palette-aware luminance editing (Figs.~\ref{fig:teaser}, \ref{fig:rgb-tone-curves}, \rev{\ref{fig:pipeline}}, \rev{\ref{fig:rollercoaster}}, \ref{fig:gallery1}), and pixel-level color constraints (Figs.~\ref{fig:teaser}, \ref{fig:gallery1}); see the supplemental for more results. All edits run in real-time, propagate consistently across novel views, and can bake back into vanilla 3DGS for use with standard viewers.

\subsection{Implementation Details and Performance}
\label{sec:implementation}
We implement our method in PyTorch with custom CUDA kernels for efficient weight splatting, and the real-time editing interface in Rust on top of Brush~\cite{brush2024}, a WebGPU-based 3DGS viewer. The interface sustains ${\sim}200$~FPS on a MacBook Pro with Apple Silicon (M5 Pro, 18-core CPU, 20-core GPU, 24 GB unified memory). Across all examples, we finetune with weight SH of degree $L = 3$, the rank-1 factorization of \S\ref{sec:low_rank} for higher-order SH, and $K = 5$ or $6$ palette colors. Each scene is finetuned for 30--50 epochs from a pretrained vanilla 3DGS checkpoint using Adam~\cite{kingma2014adam} with views shuffled per epoch, taking 40~s to 3~min on NVIDIA H200 (Table~\ref{tab:nvs_comparison}). Chromatic palette vertices $\{\mathbf{p}_i\}_{i=3}^{K}$ are initialized from a simplified convex hull~\cite{Tan:RGB2016} of the frozen renders with quantization; black and white are fixed. Hyperparameters: $\lambda_{\mathrm{pos}} = 0.5$, $\lambda_{\mathrm{compact}} = 0.001$ (ramped $5\times$ after the first $2/3$ of training), $\lambda_{\mathrm{sep}} = 1.0$ with minimum hue-gap fraction $0.3$ (\S\ref{sec:optimization}). For transparent-background scenes, we add an alpha-channel supervision term ($\lambda_{\mathrm{alpha}} = 1.0$). After $70\%$ of training we clamp barycentric supervision targets to the simplex to prevent negative splatted weights (which would invert palette edits at those pixels). We follow the evaluation protocol of \citet{barron2021mip}: hold out every 8th image as a test view, and train vanilla 3DGS baselines with uniform default settings across all scenes.
At edit time, the constraint-driven solver (\S\ref{sec:editing_solver}) runs on a single CPU thread and converges in ~8 outer iterations of damped BCD ($\alpha = 0.7$). All edits complete in 5--15~ms. 

%%%%% Ted proofread above
\subsection{Evaluation}
\begin{figure*}[t!]
\includegraphics[width=0.84\linewidth,scale=1]{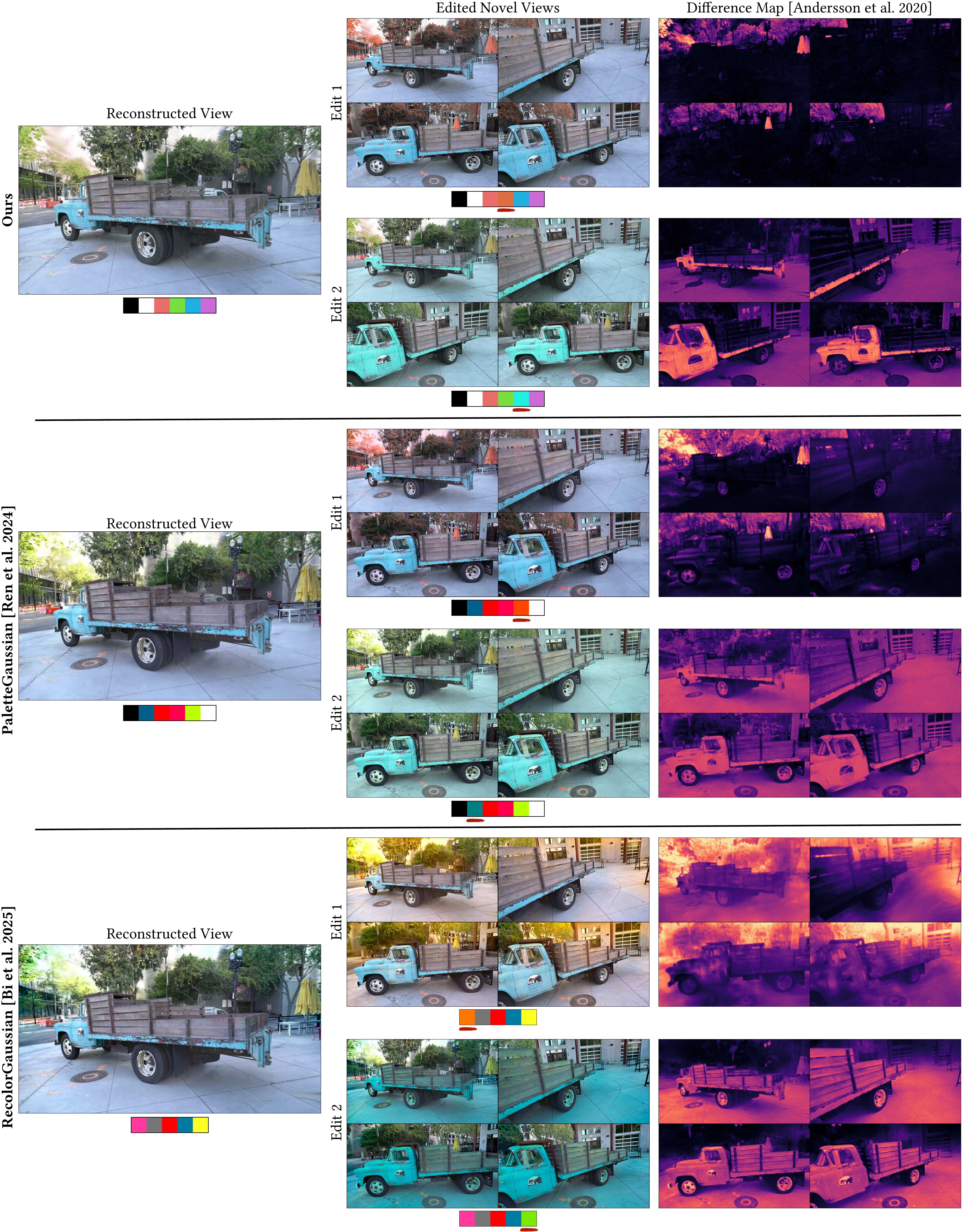}
\caption{\textbf{Sparsity comparison against primitive-space palette methods.} We compare against PaletteGaussian~\cite{PaletteGaussian} and RecolorGaussian~\cite{RecolorGaussian} on two palette edits applied to the same scene (\textit{Edit 1}: trees green to brown; \textit{Edit 2}: truck cyan to teal/purple). Difference maps~\cite{andersson2020flip} (right) show that our edits remain localized to the targeted region while the baselines bleed into unintended regions. Although the baselines' per-Gaussian weights are sparse, alpha blending during rasterization mixes contributions from many Gaussians at each pixel, destroying sparsity in view-space where edits actually happen. Our view-space color consistency loss (Sec.~\ref{sec:barycentric_supervision}) enforces sparsity at the pixel level directly. Additional comparisons are provided in the supplemental. \emph{Scenes:} \textit{Truck} from Tanks \& Temples~\cite{knapitsch2017tanks}.}
\label{fig:sparsity-compare}
\end{figure*}

We evaluate on three aspects: reconstruction fidelity to verify that editability does not degrade too much on novel-view synthesis (Table~\ref{tab:nvs_comparison}); edit locality and consistency under palette and tone-curve modifications (Fig.~\ref{fig:sparsity-compare}); and pixel-level editing against primitive-space baselines. The supplemental presents additional evaluations and comparisons (\S E), a proof of lossless palette bake-back together with finetuning time and quality for tone-curve bake-back (\S B), and an extensive investigation of a Lab-space variant (\S D, closer to~\citet{chao2023colorfulcurves}) that splats palette weights and lightness as separate SH-encoded quantities and edits in CIELAB \textit{ab}-space, which informed our choice of RGB by exposing two structural drawbacks: a non-linear Lab-to-RGB transform at edit time and the loss of closed-form palette bake-back.

We compare against our implementations of recent palette-based 3DGS methods \footnote{\rev{Code is available at \href{https://github.com/tedchao/ReparamGS-Palette/tree/main/others}{https://github.com/tedchao/ReparamGS-Palette/tree/main/others}}}, PaletteGaussian~\cite{PaletteGaussian} and RecolorGaussian~\cite{RecolorGaussian}, both of which decompose appearance in primitive space (per Gaussian). Despite per-Gaussian sparsity regularization, alpha blending during rasterization mixes contributions from many Gaussians at each pixel, so pixel weights are non-sparse in view-space and palette edits bleed into unintended regions (Fig.~\ref{fig:sparsity-compare} and supplemental \S E.2). For pixel-level editing (\S E.3), the primitive-space baselines require joint gradient descent over palette and per-Gaussian SH, taking 3.75--13.02~s per edit on NVIDIA H200 while still exhibiting color bleeding; over-regularizing the palette pushes the optimizer to compensate through the view-dependent SH originally intended for specular effects, producing an unnatural appearance (Fig.~6 in supplemental). We additionally compare against a 2D editing heuristic that edits a single rendered view with ColorfulCurves~\cite{chao2023colorfulcurves} and finetunes the per-Gaussian SH to reproduce the edit (\S E.1). Because supervision is available only on the edited view, the heuristic produces visible artifacts in novel views and adds $5.2$~s of SH finetuning per edit. In contrast, our view-space formulation achieves localized edits through a constraint-driven solve over the palette $\mathbf{P}$ and tone curves $\{f_i\}$ alone, with all Gaussian parameters fixed, propagating consistently across novel views. PaletteGaussian additionally proposes object-level editing through a second-stage training pass with segmentation masks. The masks shown in their work are coarse and lack the soft boundaries needed for natural color compositing in complex regions such as forests, as noted by~\citet{chao2023colorfulcurves}. More critically, even when segmentation succeeds, the underlying primitive-space weights are still non-sparse after alpha blending, so the same locality limitations and color bleeding persist within each segmented region.

\section{Conclusion}
\label{sec:conclusion}
Palette-based image editing follows a simple recipe: take an image, reparametrize it for editability. We present a real-time color grading approach for 3D Gaussian Splatting that applies the same principle in 3D. Our method reparametrizes a pretrained vanilla 3DGS into a view-space palette decomposition, in which each Gaussian carries weight SH that alpha-blend into per-pixel mixing weights over a small, scene-adaptive palette. We supervise the reparametrization with differentiable barycentric targets to obtain sparse, color-consistent weights, and we encode the higher-order weight SH with a rank-1 factorization that roughly halves the storage of vanilla 3DGS. We design a constraint-driven solver based on dBCD with IRLS that jointly satisfies palette, tone-curve, and pixel-level color constraints in tens of milliseconds per edit, and the edited representation bakes back into a vanilla 3DGS: palette edits in closed-form, tone-curve edits in seconds of finetuning. Extensive experiments demonstrate the effectiveness of our approach across a variety of scenes and editing scenarios.

\paragraph{Limitations and Future Work.}
Our method has several limitations that suggest directions for future work. First, our representation requires a brief finetuning stage from a pretrained vanilla 3DGS; deriving an analytical mapping from color SH to weight SH given a palette would remove this stage entirely. Second, tone-curve edits cannot be baked back in closed-form because the per-pixel luminance shift is non-linear in the rendered color, so they require a brief finetuning stage. Reducing or eliminating this stage, for instance by approximating the luminance shift with a piecewise-linear correction baked into the weight SH, is an interesting direction. Third, our method does not support semantic or object-level editing constraints. Incorporating segmentation-aware decompositions~\cite{chao2023locopalettes, aksoy2018semantic}, as demonstrated for NeRFs~\cite{chenkarf}, could enable selective recoloring of specific objects or materials while preserving our view-space sparsity advantages.

\begin{acks}
% \section*{Acknowledgements}
% None for now.
We thank our anonymous reviewers for their constructive feedback.
% Experiments Z and Y would not have been conducted without their encouragement.
This work was supported by the \grantsponsor{usnsf}{United States National Science Foundation}{http://nsf.gov/} (\grantnum{usnsf}{IIS-2402893}).
\end{acks}

\bibliographystyle{ACM-Reference-Format}
\bibliography{bib/merged,bib/gaussian}

%% https://tex.stackexchange.com/questions/45609/is-it-wrong-to-use-clearpage-instead-of-newpage
%% Clearpage eats figures
% \clearpage

% \phantom{.}

%%%%%%%% Figure Pages %%%%%%%%%%%

\clearpage

\end{document}